\newlength\smallfigwidth
\documentclass[pra,apl,10pt,twocolumn,floatfix,showpacs,amsmath,amssymb,superscriptaddress]{revtex4-1}

\usepackage{graphicx}
\def\ba{\begin{eqnarray}}
\def\ea{\end{eqnarray}}
\def\be{\begin{equation}}
\def\ee{\end{equation}}

\graphicspath{{figures-paper/}}

\begin{document}


\title{Shedding light onto topological insulator beads: perspectives for optical tweezing application}

\author{Y.G. Muller}
\email{yurimullergomes@gmail.com}
\affiliation{Centro Brasileiro de Pesquisas F\'isicas, 22290-180, Rio de Janeiro, Brazil}

\author{W.A. Moura-Melo}
\email{winder@ufv.br\\ URL: https://sites.google.com/site/wamouramelo/home}
\affiliation{Departamento de F\'isica, Universidade Federal de Vi\c cosa, 36570-000, Vi\c cosa, Brazil}

\author{J.M. Fonseca}
\email{jakson.fonseca@ufv.br}
\affiliation{Departamento de F\'isica, Universidade Federal de Vi\c cosa, 36570-000, Vi\c cosa, Brazil}

\begin{abstract}
The interaction of electromagnetic radiation with a spheric-type three-dimensional topological insulator (TI) bead is described within classical optics framework. By virtue of the topological magnetoelectric effect (TMEE) experienced by reflected and transmitted rays at the TI surface, there appears a net constant force on the spherical bead which is proportional to the fine structure constant times the incident radiation power. Such an uniform dynamics (constant acceleration) may be particularly useful for optical tweeezing techniques, for instance, to investigate a DNA strip or a membrane piece under stretching as well as to displace a tiny object by means of purely optical control.

\end{abstract}
%
%


\maketitle


\section{Introduction and Motivation}

Optical tweezing (OT) technique may be achieved whenever a highly focused laser beam is shone onto a small {\em transparent} object (say, a microsized sphere) immersed in water or other suitable medium. Whenever the sphere refraction index is greater than water one (like the widely used polysterene or silica microspheres), the refractive force balance the reflective drifting enabling bead trapping around the optical focus, as illustrated in Fig. 1. Eventually, by displacing the laser beam, or the background plate support, the sphere may be also moved. Optical tweezing appears to be a very important technique to manipulate objects sensitive to tiny forces, around {\em picoNewton} scale, namely, to study small systems, from single molecules to those around 100 $\mu$m, without direct contact. Among many other applications, OT has been shown to be a valuable tool for biophysics whenever investigating mechanical properties of proteins, membranes, DNA/RNA, and so forth\cite{reviewOT}.\\ 

Three-dimensional topological insulators (TI's, for short) comprise a class of quantum state of matter
with an insulating bulk and gapless metallic surface states protected by time reversal symmetry. Typical examples include crystals like Bi$_2$Se$_3$, Bi$_2$Te$_3$ and Sb$_2$Te$_3$. Among other peculiarities, a surface charge carrier moves in such a way that its linear momentum is always perpendicular to its spin, the so-called {\it spin-momentum locking} \cite{reviewTI}. The surface states are robust against non-magnetic impurities, whereas a magnetic disorder breaks time reversal symmetry and it open a gap at
the Dirac points. Whenever Fermi level lies within this gap, an external electromagnetic field induces Hall effect, with conductivity $\sigma_{xy}=e^2/2h$, Ref. \cite{reviewTI}. Magnetic dopant, such as $Fe$ or $Mn$, yields a gap around $E_g\backsim 50$ meV for as-grown Bi$_2$Se$_3$ crystals, as reported in Ref. \cite{chen}.\\

By virtue of these unique microscopic properties, it follows that an external electric (magnetic) field induces a magnetization (polarization) in the TI, the so-called topological magnetoelectric effect (TMEE). In addition, whenever interacting with light the topological insulator rotates the polarization plane of reflected and transmitted/refracted beams. Indeed, although the reflected ray experiences a small Faraday rotation, $\theta_F\approx \alpha \sim 1/137$ ($\alpha$ is the fine structure constant), the transmitted wave is affected by a {\em giant} Kerr effect, $\theta_K\approx \pi/2$. Such a pronounced Kerr rotation is the effect behind the appearance of a residual constant force, $\cal{\vec{F}}_{\rm Kerr}$, exerted on a small TI-made bead by a focused light. Although tiny, around some dozens of {\em femtoNewtons}, $\cal{\vec{F}}_{\rm Kerr}$ may be useful for stretching or displacing small objects with optical control, thus opening the possibilities to investigate soft matter and biophysical systems under true dynamical response.

\section{A survey on optical tweezing}

When light ray is sent to a bead, the reflection always push it along the beam by radiation pressure:
\begin{equation} \label{Fdrift}
\vec{F}_{RP}=\frac{n_m}{c}\,\int_{\cal A} \vec{S} d{\cal A}= \frac{2n_m}{c} P_{\rm beam} \hat{n}_{\rm beam}\,,  
\end{equation}
$n_m$ is the refractive index of the medium surrounding the particle (OT techniques frequently utilizes de-ionized water, $n_m\approx 1.343$), $c\approx 3\times 10^8\, {\rm m/s}$ is the vacuum light speed, and $P_{\rm beam}$ is the laser beam power, around {\em miliWatts} in typical experiments. Therefore, we are talking about very tiny forces, ranging at {\em picoNewton} scale, $\sim 10^{-12}\,{\rm N}$. Alternatively, this force may be computed by $\vec{F}_{RP}=(n_m/c)\, \sigma \Big<\vec{S}\Big>_t$, where $\sigma$ is the sphere cross-section and $\Big<\vec{S}\Big>_t$ is the time-average of the Poynting vector\cite{reviewOT}.\\

In turn, 'transparency' of the bead provides restoring forces comes from light refraction. Two distinct limiting cases, concerning the ratio between the particle size/radius, $a$, and radiation wavelength, $\lambda$, are in order:\\

{\bf A)} Rayleigh regime: $\lambda>>a$
\begin{equation}
\vec{F}_{1}=\frac{K-1}{K+2} a^3 \vec{\nabla}(|\vec{E}|^2)\,,
\end{equation}
which is the so-called gradient force, pointing towards $\vec{E}$ is more strength, say, the optical focus (Fig. 1). $K=\epsilon/\epsilon_m$ is the ratio between particle and medium electric permittivities.\\

{\bf B)} Geometrical Optics regime: $\lambda<<a$\\
Now, the net restoring force coming from all multiple refractions occurring inside the bead may be estimated, considering a Gaussian beam profile, to read like below\cite{reviewOT}:
\begin{equation}
\vec{F}_2= \frac{n_m}{c}[{\rm Re}(Q_t) \hat{z} + {\rm Im}(Q_t)\hat{y}]P_{\rm beam}\,,
\end{equation}
where $Q_t$ is defined by:
\begin{equation}\label{Qt}
Q_t\equiv 1 + R e^{2i\theta_i} -T^2\frac{e^{2i(\theta_i-\theta_t)}}{1 + R e^{-2i\theta_t}}\,,
\end{equation}
\begin{equation}
R(\theta_i,\theta_t)=\frac12 \left[\frac{\sin(\theta_i-\theta_t)}{\sin(\theta_i+\theta_t)}\right]^2 + \frac12 \left[\frac{\tan(\theta_i-\theta_t)}{\tan(\theta_i+\theta_t)}\right]^2\,.\nonumber
\end{equation}
$\theta_i$ and $\theta_t$ are the incident and refracted angles, Fig. 1. Recall that Reflectivity + Transmittivity equals unity, $R+T=1$.\\
It should be emphasized that for an arbitrary ratio, $\lambda/a$, Mie-Debye treatment taking into account spherical aberration is necessary.\\
In words, the two optical forces act in opposite ways: the radiation pressure always push the object tending to drift it away, whereas the restoring force, provided by light refraction, pulls the bead towards the optical focus. Whenever the dielectric particle refraction index is greater than that of the surrounding medium, refractive force overcomes radiation pressure in such a way that the optical focus comes to be the stable point, around which the bead is trapped.  Further details may be found in Ref. \cite{reviewOT}.

\begin{figure}[!h]\label{OT-forces}
 \begin{center}
 \includegraphics[width=6cm]{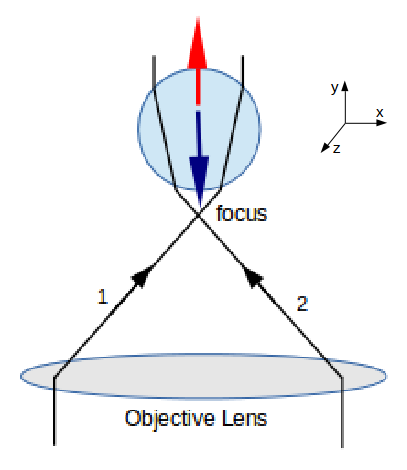}\vskip .1cm
 \includegraphics[width=8cm]{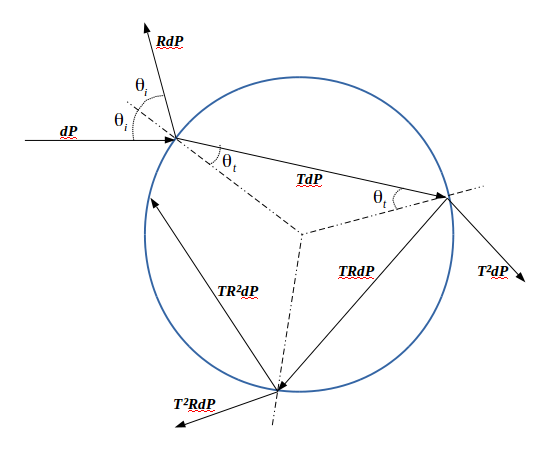}
 \end{center}
{\caption{\small Basic forces involved in optical tweezing. (Top panel): A drift force coming from reflected rays push the bead away along the incident beam direction, red arrow. Refractions, including those coming from multiple internal processes (panel below), combine to give a net restoring force towards the focus, blue arrow. (Bottom panel): shows splitting rays coming about from multiple reflection and refraction of a incident ray, ${dP}$, at external and internal spherical bead surfaces. Remaining labels account for the number of times a given ray has been reflected, ${R}$, or transmitted/refracted, ${T}$. For instance, ${ TRdP}$, means that ${dP}$, has been transmitted once and reflected a unique time. According to the geometrical optics regime applied to optical tweezing one need to compute all the multiple rays contributions to obtain the net refractive and reflective forces acting onto the bead. $(\theta_i,\theta_t)$ account for the (incident, refracted) angles of the primary rays.}} 
\end{figure}

\section{{Light interacting with TI: magnetoelectric effect}}
A magnetic field perpendicularly applied to the surface of a topological insulator yields time reversal symmetry breaking there, providing energy gap for such states. Whenever the Fermi level lies in the gap, eventually spin quantum Hall effect takes place which may be described by a topological magnetoelectric effect at the long-wavelength regime, $\alpha\Theta$-term appearing in action below:
\begin{equation}
{\cal S}=-\frac12 \int d^4 x \left[\epsilon |\vec{E}|^2 -\frac{1}{\mu} |\vec{B}|^2 +\alpha\frac{\Theta}{2\pi} \vec{E}\cdot \vec{B}\right]\,,
\end{equation}
where $\epsilon$ and $\mu$ are the permittivity and permeability of TI-material, while $\Theta=0$ (or ${\rm or}\, \pi$) for an ordinary (or topological) insulator. Maxwell equations keep their forms, provided that we take into account the following constitutive relations:
\begin{equation}
\vec{D}=\epsilon \vec{E} +\alpha \frac{\Theta}{\pi}\vec{B}\quad {\rm and} \quad \vec{H}=\frac{\vec{B}}{\mu} - \alpha \frac{\Theta}{\pi}\vec{E}\,.
\end{equation}
Physically, an applied electric (and/or magnetic) field induces an extra 'topological-like' {\em magnetization} (and/or {\em polarization}), $\vec{P}=(\alpha\Theta/\pi)\vec{B}$ and/or $\vec{M}=- (\alpha\Theta/\pi)\vec{E}$. The scenario may be understood as induced sources on the TI surface exhibiting Hall effect \cite{reviewTI}, say, $\sigma_H=\hat{n}\cdot \vec{P}= \alpha\hat{n}\cdot \vec{B}$ and $\vec{K}_H=\hat{n} \times \vec{M}= -\alpha\hat{n}\times \vec{E}$, where $\sigma_H$ and $\vec{K}_H$ are the Hall charge and current densities induced on the TI surface as a response to the applied fields.\\

Whenever light comes into the game, the non-trivial boundary conditions imposed by the Hall sources on the normal electric, $\epsilon_2 (E_2)_N-\epsilon_1 (E_1)_N = \alpha{(B_1)_N}$, and transverse magnetic fields, $(B_2)_T/\mu_2 -(B_1)_T/\mu_1= \alpha (E_1)_T$, along with the usual ones for $\vec{E}_T$ and $\vec{B}_N$, yield topological Kerr and Faraday effects given by\cite{hugeKerr}:
\begin{equation}
\theta_K \approx \arctan (1/\alpha) \approx \pi/2 \quad {\rm and} \quad \theta_F\approx \arctan(\alpha)\approx \alpha\,.
\end{equation}
Although the refracted/transmitted beam is almost insensitive to these Hall sources, once $\theta_F\approx \alpha\sim 1/137 <<1$, the reflected rays experience a huge change in their polarization direction, $\theta_K\approx \pi/2$ (note that $R$ and $T$ are affected only by very small contributions, say, at second order in $\alpha$: $(R,T)= (R_{\rm usual}, T_{\rm usual}) + {\cal O}(\alpha^2)$). Such a giant Kerr effect, mainly acting on the secondary rays producing multiple reflections at the interface inside the sphere, open us the possibility of using light to move TI-made beads by means of an extra optical force coming from topological grounds, as below.\\
 
\section{{Topological force on a TI-made tiny sphere}}

Our main result comes about from the simple fact that the huge Kerr effect yields a non-vanishing radiation pressure on a TI-made sphere which is {\em perpendicular to the incidence plane}, $yz$. Such a force presents a leading term linear in $\alpha$ and it is directed along $x$, as illustrated in Fig. 2, and may readily obtained as follows:
\begin{equation}\label{FKerr}
\vec{\cal F}_{\rm Kerr}= \frac{n_m}{c} \sigma \Big<\vec{S}_{\rm Kerr}\Big>_t= \alpha\,\Omega |\vec{F}_{\rm RP}| \, \hat{x} \,,
\end{equation}
where $\sigma$ is the TI-sphere cross-section and $\Big<\vec{S}_{\rm Kerr}\Big>_t$ is the time-average of the Poynting vector associated to the Kerr reflected radiation; We have also used, $|\vec{F}_{\rm RP}|= 2\frac{n_m}{c} P_{\rm beam}$,  Eq. (1). The quantity $\Omega=\Omega(n_m,n,\theta_i, \theta_t)$ accounts for the summation over multiple reflections and refractions (ray optics framework, $\lambda<<a$). [Whenever working within Rayleigh regime, $\lambda>>a$, one realizes that $\vec{\cal F}_{\rm Kerr}$ identically vanishes once upper hemisphere contribution exactly cancels that coming from below the sphere equator. As a net result, the bead is expected to only self-rotates at this regime, without displacing elsewhere]. It is a length expression depending on the refractive indexes and on the incident and transmitted angles, similarly to the factor $Q_t$, Eq, (\ref{Qt}), and it is evaluated in the Appendix. It may be also faced as an effective cross-section (form-factor) of the TI-sphere whenever experiencing Kerr rotation. Fig. 2 shows the behavior and magnitude of $\Omega$ as function of $\theta_i$.\\
Note that $|\vec{\cal F}_{\rm Kerr}|$ goes around some dozens of {\em femtoNewton}, say, $\sim 10^{-2} - 10^{-3}$ times radiation pressure magnitude $|\vec{F}_{RP}|$. Therefore, it is small to jeopardize laser focalization on the colloidal particles. On the other hand, it may be useful to move very small particles, like single molecules and molecular motors, with constant acceleration perpendicularly to the polarization plane, say, $x$-axis, as depicted in Fig. 2. It may also be useful to investigate a number of dynamical properties of polymers, membranes etc under controlled mechanical stretching provided by a constant force.\\

Furthermore, let us recall that typically used dielectric spheres have around a few microns, say, $5\mu{\rm m}$, leading to mass about $\sim 10^{-13}\,{rm Kg}$ (density around $1-1.5\times 10^3 \,{\rm Kg/m^3}$, for polystyrene and mica). Therefore, radiation pressure or gradient force would input high acceleration to the beads, $\sim 10 \,{\rm m/s^2}$, if they acted isolated. The topological force brought about here produces a much smaller acceleration: taking TI-sphere with $r=5\mu{\rm m}$ and density $\sim 7\times 10^3 \,{\rm Kg/m^3}$ (value for $Bi_2 Se_3$), then $\vec{\cal F}_{\rm Kerr}$ yields acceleration around $|\vec{a_{Kerr}}|\sim 10^{-2} - 10^{-1}\, {\rm m/s^2}$. Such an acceleration is capable of producing high mobility to the TI-beads, making them to displace thousands of their size in just 1 second.

\begin{figure}[!h]
 \begin{center}
 \includegraphics[width=6cm]{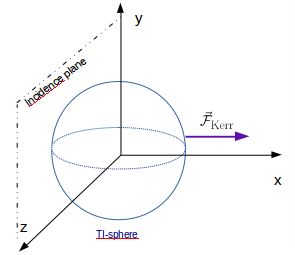}\vskip .1cm
 \includegraphics[width=8cm]{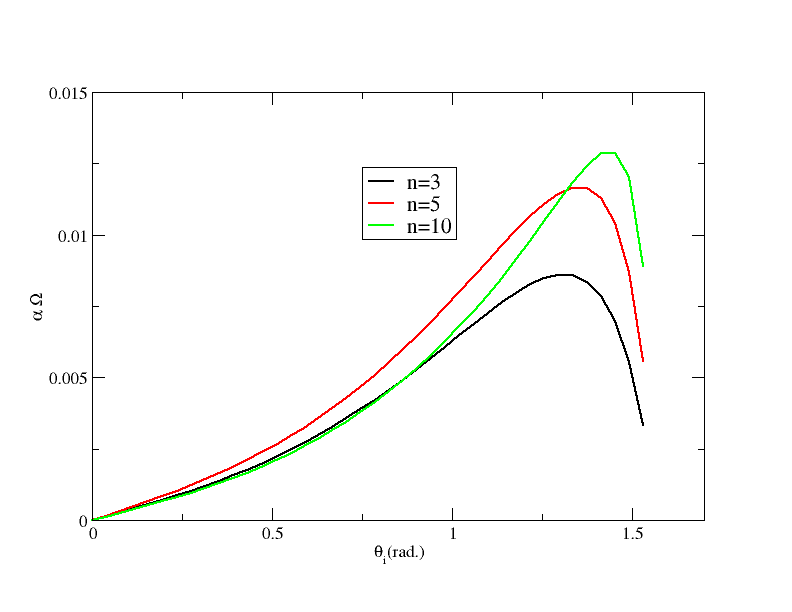}
 \end{center}\label{F-Kerr-grafico}
{\caption{\small Orientation and magnitude of topological-like Kerr force. (Top panel): Although tiny, $\vec{\cal F}_{\rm Kerr}\sim 10-10^2 \,{\rm fN}$, it may be used to push very small bodies attached to TI-bead along the direction perpendicular to the incidence plane. (Bottom panel): Multiple reflections/refractions processes, relevant in the geometric optics regime, $a>>\lambda$, is responsible for the appearance of such a topological force. The relative magnitude of Kerr and radiation pressure forces, $|\vec{\cal F}_{\rm Kerr}|/|\vec{F}_{RP}|=\alpha\Omega$, is plotted against the incident angle, $\theta_i$, for a number of typical TI refraction indexes, $n=3, 5,\,{\rm and} \,10$ (we have taken $n_m=1.343$ for de-ionized water). Further details upon $\Omega$ may be found in the Appendix.}} 
\end{figure}

\section{{Conclusions and Prospects\\}}
Small TI-made beads experience a tiny residual force perpendicular to the incident plane, $\vec{\cal F}_{\rm Kerr}$, whenever interacting with light which may be relevant for molecular and biophysical studies, namely, for displacing and/or stretching tiny particles, like molecules and small protein pieces, under a purely optically controlled motion.\\

From the experimental point of view, the measurement of $\vec{\cal F}_{\rm Kerr}$ may be challenging. At least two aspects must be emphasized:\\
First, TI-made beads surface should be submitted to a perpendicular magnetic field in order to open an energy gap responsible for the appearance of the TMME and hence the giant Kerr rotation. The best way to achieve this condition is by growing TI-beads around an even smaller sphere/bead of ferromagnetic material.\\ Second, Kerr force has a tiny magnitude. For instance, whether and how it can overwhelm viscous forces in colloidal particles systems is one of the key points towards its detection. For that, TI-made beads with precise size and mass along with suitable viscous liquid are in order. For instance, if we intend to use Casimir-type forces\cite{Casimir-EPL}, ranging at the same scale of $\vec{\cal F}_{\rm Kerr}$, $\sim 10-10^2$ fN, to probe it. An key feature of the Kerr force relies on its constance, in magnitude and orientation, rendering a relatively clear way to distinguish its effects (say, acceleration) from those associated to viscous and other medium-attributed forces, generally expected to be almost anisotropic and working randomly on the spherical bead.\\

\centerline{\bf Acknowledgements\\}
The authors express their gratitude to M.S. Rocha for fruitful discussions regarding Optical Tweezing technique and its applications to Biophysical systems. They also thank CAPES, CNPq, and FAPEMIG (Brazilian agencies) for financial support.\\

\centerline{\bf Appendix\\}

In order to take into account all the contributions coming from multiple reflections taking place on the internal surface of the TI-sphere, we must sum the Kerr effect over all internally reflected rays, denoted by ${TdP}$, ${TRdP}$, ${TR^2dP}$, etc, as depicted in Fig. 1 (bottom panel). There, ${dP}$ represents the (infinitesimal) power of an arbitrary incident beam, while coefficients ${R}$ and $T$  account for the {\em Reflectivity} and {\em Transmittivity} of the bead. Therefore, a secondary ray, for instance, ${TdP}$ has a power $T$ times that carried by the incident ray, and so forth. At leading order, multiple internal reflections (MIR) yield:
\begin{eqnarray}
&\Big<d\vec{S}_{\rm MIR}\Big>_t &= \alpha dP \left[ f_T T + f_T {TR} +  f_T {TR^2} + \ldots \right]\, \hat{x}\nonumber\\
& & = \alpha dP \left[ f_T T \sum_{n=0}^\infty { R}^n \right]\, \hat{x}\equiv \alpha dP  \, \Omega \, \hat{x} \label{EqOmega}
\end{eqnarray}
where $f_T=f_T(\theta_i, n_m, n_b)$ is given, at zero-th order in $\alpha$, by:
$$f_T= \frac{4 n^3_m (\frac{n^2}{n^2_m} -1) \sin(2\theta_i)}{(n^2_m + n^2 + n_m n\, \xi_+)} \left[1+ n \xi_- +\frac{n}{2n_m} \xi_+ \right]\,,$$
with $\xi_\pm=\frac{\cos^2\theta_i \pm \cos^2\theta_t}{\cos\theta_i\,\cos\theta_t}$. Therefore, $\Omega=\Omega(\theta_i, n_m, n)$ is but nothing an extra factor taking into account the effects of multiple internal reflections to the topological Kerr force, $\vec{\cal F}_{\rm Kerr}$. Integrating Eq. (\ref{EqOmega}) overall the power beam and summing up the Kerr effect coming from the primary reflected ray, ${RdP}$ in Fig. 1, we finally obtain the result of Eq. (\ref{FKerr}). Further details on how to compute multiple internally reflected and refracted processes, may be found in Refs. \cite{reviewOT}, and in other references cited therein.\\

\end{document}